\documentclass[conference]{IEEEtran}
\IEEEoverridecommandlockouts
\usepackage{cite}
\usepackage{amsmath,amssymb,amsfonts}
\usepackage{algorithmic}
\usepackage{graphicx}
\usepackage{textcomp}
\usepackage{xcolor}
\usepackage{hyperref}
\hypersetup{hidelinks}

\def\BibTeX{{\rm B\kern-.05em{\sc i\kern-.025em b}\kern-.08em
    T\kern-.1667em\lower.7ex\hbox{E}\kern-.125emX}}
\begin{document}

\title{Big Data Architecture for Large Organizations\\
}

\author{\IEEEauthorblockN{1\textsuperscript{st} Fathima Nuzla Ismail*}
\IEEEauthorblockA{\textit{Dept. of Mathematics} \\
\textit{State University of New York at Buffalo}\\
USA \\
fathima.nuzla.ismail@gmail.com}
~\\
\and
\IEEEauthorblockN{2\textsuperscript{nd} Abira Sengupta}
\IEEEauthorblockA{\textit{School of Computing} \\
\textit{University of Otago}\\
Dunedin, New Zealand\\
sengupta.abira0609@gmail.com}
%*Corresponding author
~\\
\and
\IEEEauthorblockN{3\textsuperscript{rd} Shanika Amarasoma}
\IEEEauthorblockA{
\textit{Independent Researcher}\\
Sri Lanka\\
shanika.amarasoma@gmail.com}
~\\

}

\maketitle

\begin{abstract}
The exponential growth of big data has transformed how large organisations leverage information to drive innovation, optimise processes, and maintain competitive advantages. However, managing and extracting insights from vast, heterogeneous data sources requires a scalable, secure, and well-integrated big data architecture. This paper proposes a comprehensive big data framework that aligns with organisational objectives while ensuring flexibility, scalability, and governance. The architecture encompasses multiple layers, including data ingestion, transformation, storage, analytics, machine learning, and security, incorporating emerging technologies such as Generative AI (GenAI) and low-code machine learning. Cloud-based implementations across Google Cloud, AWS, and Microsoft Azure are analysed, highlighting their tools and capabilities. Additionally, this study explores advancements in big data architecture, including AI-driven automation, data mesh, and Data Ocean paradigms. By establishing a structured, adaptable framework, this research provides a foundational blueprint for large organisations to harness big data as a strategic asset effectively.
\end{abstract}


\section{Introduction}

The rapid growth of data has changed how big businesses make
decisions and compete in a market driven by data. Big data, identified by its volume, speed, and variety of formats, adds complexity and offers huge potential\cite{Lutfiani:2024, Speckhard:2025}. Big data can be used effectively by large organizations to encourage innovation, streamline processes, and quickly develop. However, a complex architecture that takes into account scalability, integration, governance and security is necessary for managing and drawing conclusions from such vast amounts of data \cite{Sivarajah:2024, Theodorakopoulos:2024}.

A structured approach to addressing these challenges is implementing a significant big data architecture, which serves as
a guiding framework for organisations looking to optimize their
data-driven strategies. Data ingestion, processing, storage, and analysis are all covered in this blueprint, which also includes compliance with
privacy, data quality, and regulatory standards \cite{ismail2019evaluating, Theodorakopoulos:2024, singh:2024}.

In order to prioritise scalability, flexibility, and alignment with
organisational goals, this paper suggests a big data architecture for
large organisations. The blueprint supports a variety of data types
from multiple sources by integrating data engineering frameworks
and advanced analytics \cite{Frick:2024}.

The study looks at best practices in big data architecture, cloud-based infrastructure, machine learning integration, and data ecosystems. Additionally, it discusses how vital data governance is to maintaining data security and quality. This paper aims to provide large organisations with a foundational framework to handle the demands of big data management by offering a flexible and scalable big data blueprint, establishing data as a strategic asset  \cite{Abdujapparova:2024}.

%%%%%%%%%%%%%%%%%%%%%%%%%%%%%%%%%%%%%
\section{Methodology}

This methodology describes a systematic method for creating a big
data architecture suited to most businesses' requirements.
Five steps comprise the development process: requirements collection,
architecture design, case study validation, framework implementation, and continuous optimisation. In order to satisfy organisational
standards and requirements, each stage integrates quantitative and
qualitative analyses while addressing fundamental scalability, integration, governance, and security issues. This paper focuses on
the designing architecture stage, as it is the crucial step in
designing IT blueprints in large organisations.

%%%%%%%%%%%%%%%%%%%%%%%%%%%%%%%%%%%%%%%%%%%%%%%%%%%%%%%
\subsection{Requirements Analysis}

The first step is thoroughly analysing the organisation's technology
requirements, focusing on data sources, volume,
diversity, processing speed, and intended data usage. Stakeholder
surveys and interviews are conducted during this phase, and the
current data infrastructure is evaluated. Scalability, real-time processing, privacy, security, and compliance are given particular attention because they are crucial in big data environments \cite{Theodorakopoulos:2024, Sivarajah:2024}.

%%%%%%%%%%%%%%%%%%%%%%%%%%%%%%%%%%%%%%%%%%%%%%%
%\vspace{-5pt} % Before the figure
\begin{figure*}[!htbp]
    \centering
   %\vspace{-10pt} % Adjust this value as necessary
    \includegraphics[width=0.8 \textwidth]{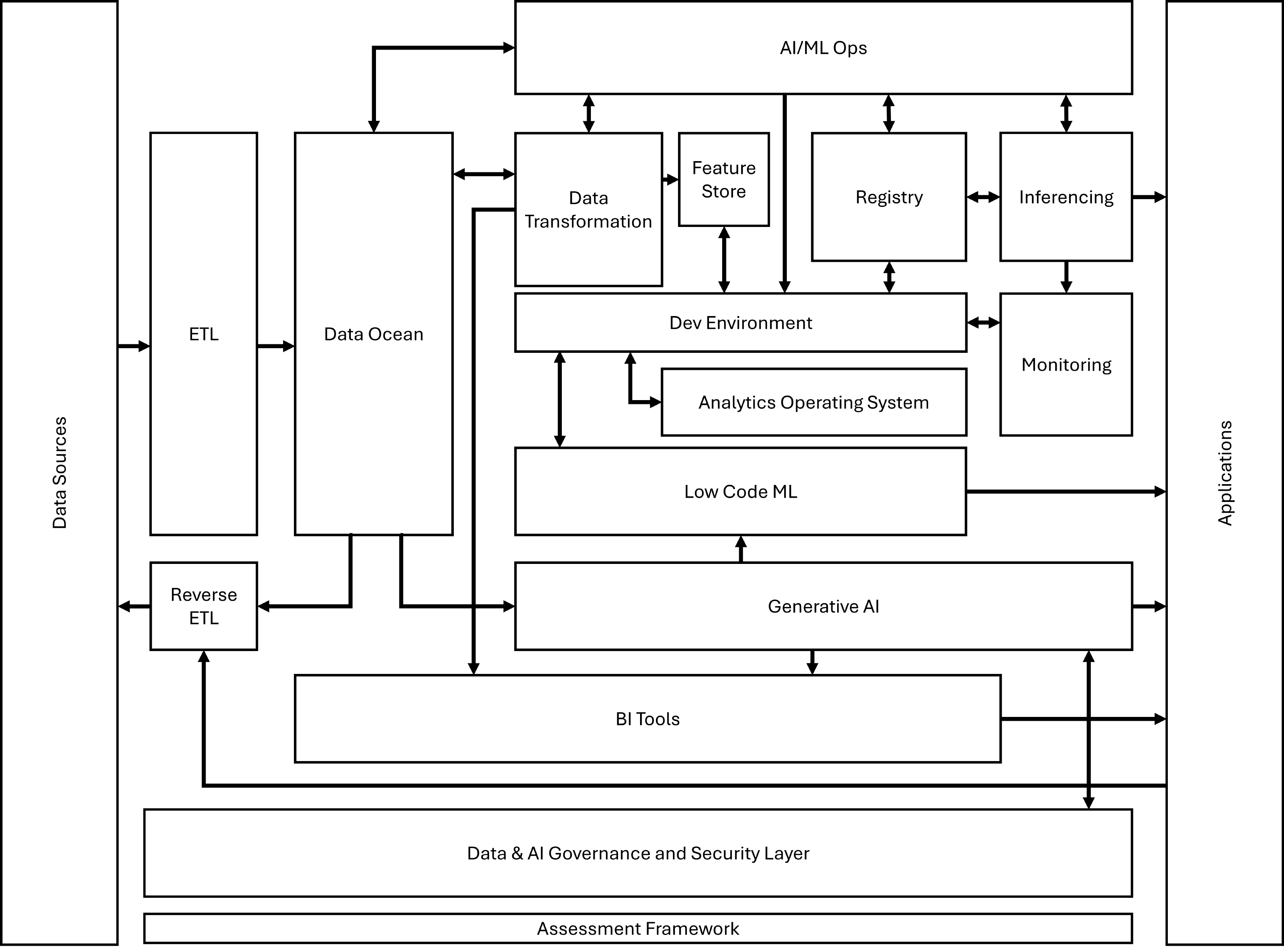} % Adjust width as needed
    %\captionsetup{ width=0.7\textwidth} % Shift caption left
    \caption{High-Level Functional Diagram}
    \label{fig:fig-1}
\end{figure*}

%%%%%%%%%%%%%%%%%%%%%%%%%%%%%%%%%%%%%%%%%%%%%%%

%\vspace{-5pt} % Before the figure
\begin{figure*}[!htbp]
    \centering
   %\vspace{-10pt} % Adjust this value as necessary
    \includegraphics[width=0.95 \textwidth]{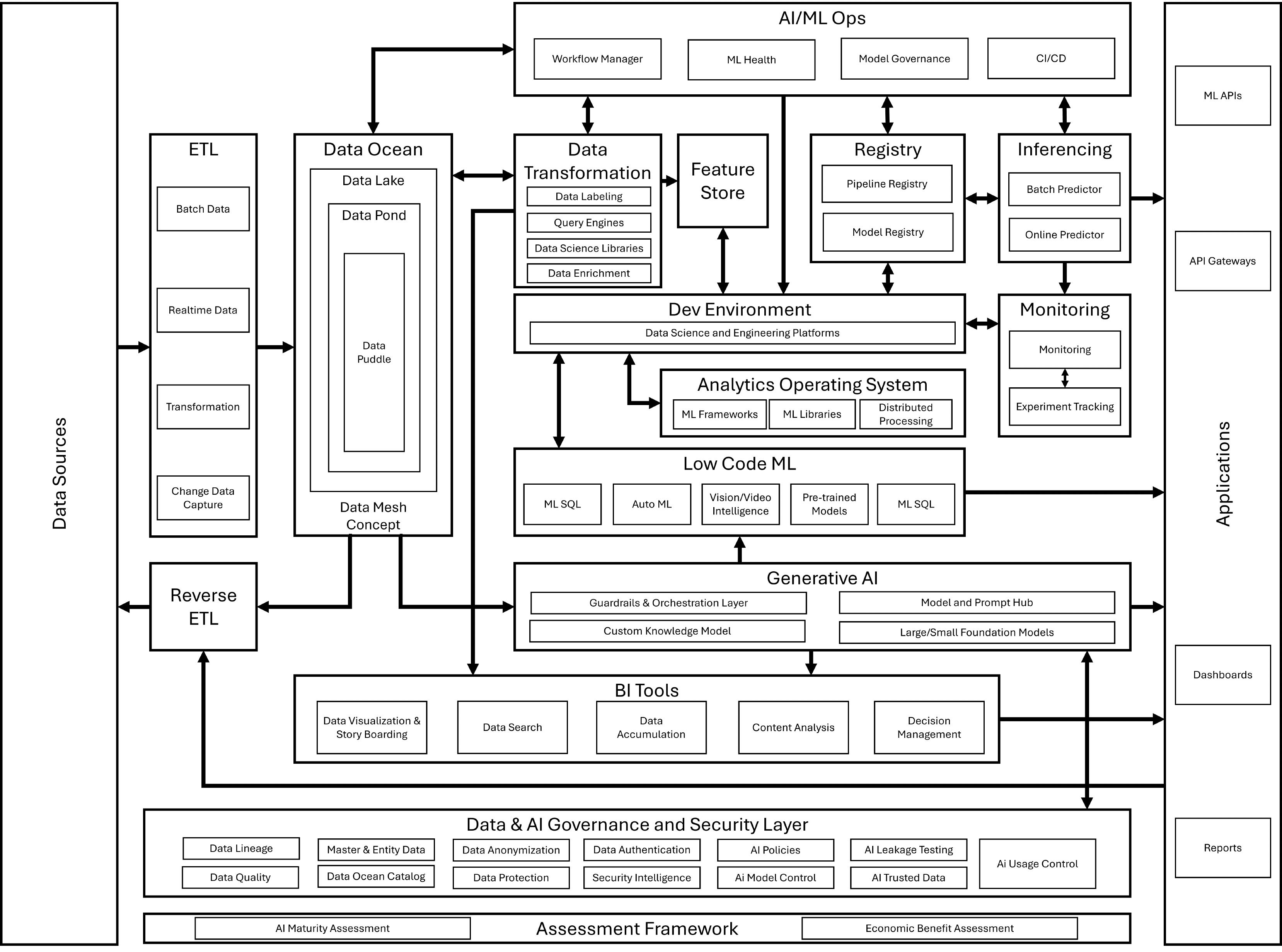} % Adjust width as needed
    %\captionsetup{ width=0.7\textwidth} % Shift caption left
    \caption{Detailed Functional Diagram}
    \label{fig:fig-1a}
\end{figure*}

%%%%%%%%%%%%%%%%%%%%%%%%%%%%%%%%%%%%%%%%%%%%%%%

\subsection{Architectural Design}
Based on requirements, a high-level functional blueprint for the architecture is developed, consisting of the following data layers\footnote{In IT architecture, layers refer to an environment that can be autonomously managed and maintained, tested, and changed independently. This allows applications to be managed in the same environment. \cite{rana2022high}} and its key components (Figure~\ref{fig:fig-1}):

%%%%%%%%%%%%%%%%%%%%%%%%%%%%%%%%%%%%%%%%%%%%%%%%%%%%%%%

\begin{itemize}
    \item \textbf{Data Source layer:} This layer accepts structured, semi-structured, and unstructured data from various sources, using batch processing and real-time streaming for flexibility. Databases, IoT devices, social media, and logs are examples of data sources \cite{Theodorakopoulos:2024}.

    \item \textbf{The Extraction, Transformation, and Load (ETL) layer consists of three key stages:} Extracting data from one or more input sources, cleaning and modifying it as needed, and then putting it into an output data container. Data aggregation from various sources is made easier by this process, which also permits output to one or more destinations. It uses cloud and distributed storage techniques to effectively handle massive amounts of data. Additionally, reverse ETL is incorporated, where data is processed within a data warehouse and then transferred to a third-party system for further action.

    \item \textbf{Data Ocean layer:} The simple premise of data mesh is that business domains should be able to define, access, and control their data products. Data is considered a product and distributed among the domains/owners.

    \item \textbf{Data Transformation layer:} This layer is intended to facilitate both batch and real-time processing such as data transformation, cleaning, and aggregation while maintaining data governance and quality \cite{Frick:2024}.

    \item \textbf{Analytics and Machine Learning layer:} This layer uses frameworks to enable business intelligence, machine learning, and predictive analytics. Large-scale analytics \cite{Demirbaga2024F} allows businesses to use data insights to inform strategic decisions. The layers were further divided into Generative Artificial Intelligence (GenAI), Business Intelligence Tools (BI tools), and low-code machine learning.
    
    \item \textbf{Low-Code Machine Learning layer:} The use of low-code machine learning (ML) solutions is growing \cite{raghavendran2023low}. Exploiting Data Science and Machine Learning modelling is now less expensive and requires fewer skills thanks to this layer. ML Structured Query Language SQL, which can feature engineer, train, test, and deploy ML models, is a component of this layer. Iterative and time-consuming ML modelling tasks are reduced with the aid of Automated Machine Learning (autoML). Particularly for large-scale ML modelling tasks like natural language processing, pre-trained models are increasingly required. The pre-trained models are typically adjusted and scaled to meet the needs of the user. In addition, trained models are frequently saved for later use in other projects.
    
    \item \textbf{AI Ops:}
    Artificial Intelligence for IT Operations (AI Ops) serves as a model registry, a central location where model developers can post models that are ready for production for convenient access. An overview of successful and unsuccessful deployments is provided by the pipeline registry, which also maintains the deployment pipelines.

    % \item \textbf{Generative AI (GenAI):}
    % Generative AI protects AI models' safety, accuracy, and fairness while managing and deploying them \cite{bandi2023power}. When creating and implementing generative AI models, model and prompt hubs promote efficiency, reproducibility, and teamwork. Generative AI is a specialised model that has been trained on particular data to produce text, translate languages, create original content, and provide end users with informative, domain-specific answers to their questions. Large foundation models are strong AI models that can be used for a variety of tasks, such as creating original content, translating languages, and producing text.

    \item \textbf{Data governance and security:}
    AI policies serve as guidelines for the responsible, ethical, and reliable development and application of AI. Model control is the process of overseeing and managing the creation, application, and deployment of AI models to guarantee their responsibility, safety, and dependability. To ensure that AI models are used responsibly, ethically, and in keeping with the law, usage control refers to the procedures and policies that regulate their use within an organisation.
    
\end{itemize}

%%%%%%%%%%%%%%%%%%%%%%%%%%%%%%%%%%%%%%%%%%%%%%%%%%%%%%%
Additional layers, such as feature stores for automating and reusing machine learning models, are added in addition to the primary layers based on business needs. Communication between various systems, apps, and users is made easier by interference. For effective data exchange, it supports middleware services, standardises data flow, and integrates internal and external systems. Additionally, we have built distinct layers for campaign management and third-party apps. A generic detailed functional architecture is presented in the (Figure~\ref{fig:fig-1a}).
%%%%%%%%%%%%%%%%%%%%%%%%%%%%%%%%%%%%%%%%%%%%%%%%%%%%%%%%
%\vspace{-8pt} 

\subsection{Validation and Case Study}

To simulate real-world conditions, the architecture is tested with a case study from a large organisation. To evaluate the efficacy of the architecture, performance metrics are evaluated, including scalability, model accuracy, storage efficiency, and data processing speed. Surveys and interviews are used to gather input from users, data engineers, and data scientists in order to guide future developments \cite{Abdujapparova:2024}.

%%%%%%%%%%%%%%%%%%%%%%%%%%%%%%%%%%%%%%%%%%%%%%%%%%%%%%%%
%\vspace{-8pt} 

\subsection{Assessment Framework}

Continuous monitoring is done to evaluate performance and make sure governance policies are being followed after the architecture is deployed. The knowledge gathered from this monitoring informs continuous changes, creating a feedback loop that maintains the architecture in line with evolving organisational requirements and technological advancements \cite{Frick:2024}. This process provides an organised approach to building a big data architecture that is secure, flexible, and scalable. Large organisations can efficiently utilise their data assets while maintaining adherence to industry standards by coordinating the design phases with organisational goals and industry best practices.

%%%%%%%%%%%%%%%%%%%%%%%%%%%%%%%%%%%%%%%%%%%%%%%%%%%%%%%%
%\vspace{-10pt}

\subsection{Continuous Monitoring and Feedback Mechanism}

Before the results could impact businesses, many tools had been introduced for model monitoring, mostly model drift. In this big data architecture, experiment tracking tools also aim to provide a comprehensive overview of the model metrics and artefacts in one location.

The majority of organisations use assessment frameworks to confirm that the benefits outweigh the costs of implementing the current Big Data platforms, ML models, pipelines, and ETLs, as well as that they meet the most recent standards. This function can be customised according to the organisation's currently deployed frameworks. Several assessment frameworks can be incorporated: Project, Data Quality, and Code Quality Assessments.

%%%%%%%%%%%%%%%%%%%%%%%%%%%%%%%%%%%%%%%%%%%%%%%%%%%%%%%%
%\vspace{-10pt}

\section{Results}

The architecture diagrams\footnote{\url{https://github.com/GenomicAI/BigData}} highlight different cloud architectures within detailed tools and components that can be used in each data layer.
%The various cloud architectures are depicted in the following figures (Figure~\ref{fig:fig-2}, \ref{fig:fig-3}, \ref{fig:fig-4}) along with comprehensive tools and components that are applicable to each data layer.

%%%%%%%%%%%%%%%%%%%%%%%%%%%%%%%%%%%%%%%%%%%%%%%%%%%%%%%%

%\begin{figure}[ptbh] 
%\includegraphics[scale=0.38]{ACM-fig-2.pdf}
%\caption{} \label{fig:fig-2}
%\end{figure}

%\vspace{-5pt} % Before the figure

% \begin{figure*}[!htbp]
%     \centering
%    %\vspace{-10pt} % Adjust this value as necessary
%     \includegraphics[width=1\textwidth]{GCP.pdf} % Adjust width as needed
%     %\captionsetup{ width=0.7\textwidth} % Shift caption left
%     \caption{Google Cloud environment}
%     \label{fig:fig-2}
% \end{figure*}

%%%%%%%%%%%%%%%%%%%%%%%%%%%%%%%%%%%%%%%%%%%%%%%%%%%%%%%%

\subsection{Google Cloud environment}

Google Cloud tool resources include official documentation, technical reports, and research articles on cloud computing platforms\footnote{https://cloud.google.com/products}. The resources listed below will be especially helpful.

%%%%%%%%%%%%%%%%%%%%%%%%%%%%%%%%%%%%%%%%%%%%%%%%%%%%%%%%

\subsubsection{\textbf{Google Cloud Documentation}} 

Google Cloud Documentation provides detailed descriptions of each tool and its functionalities, as well as best practices for implementation and management.

%%%%%%%%%%%%%%%%%%%%%%%%%%%%%%%%%%%%%%%%%%%%%%%%%%%%%%%%

\subsubsection{\textbf{Cloud Service Analysis}} 

The technical architecture and use cases of major cloud platforms are discussed, including key Google Cloud services such as \textbf{BigQuery}, \textbf{Compute Engine}, and \textbf{Cloud Storage}.

%%%%%%%%%%%%%%%%%%%%%%%%%%%%%%%%%%%%%%%%%%%%%%%%%%%%%%%%

\begin{itemize}
    \item  \textbf{Machine Learning and Data Analytics on Google Cloud:} Machine Learning and Data Analytics on Google Cloud examines Google Cloud's tools for machine learning (\textbf{Vertex AI}) and big data (\textbf{BigQuery}), with a focus on analytics capabilities.

    \item \textbf{Cloud Security and Networking:} On Google Cloud, review the security protocols, network configurations, and identity management.
    
\end{itemize}

%%%%%%%%%%%%%%%%%%%%%%%%%%%%%%%%%%%%%%%%%%%%%%%%%%%%%%%%
%\vspace{-5pt}

A comprehensive understanding of Google Cloud's offerings and architecture is supported by the insights each of these sources offers into the technical aspects, use cases, and features of its suite.

%%%%%%%%%%%%%%%%%%%%%%%%%%%%%%%%%%%%%%%%%%%%%%%%%%%%%%%%
\subsection{Amazon Web Services (AWS) environment}

The following sources highlight some of AWS's key features for users: cloud computing, storage, analytics, machine learning, and security\footnote{https://docs.aws.amazon.com}.
 
 %\vspace{5pt}

%%%%%%%%%%%%%%%%%%%%%%%%%%%%%%%%%%%%%%%%%%%%%%%%%%%%%%%%

% \begin{figure*}[!htbp]
%     \centering
%     \includegraphics[width=1\textwidth]{AWS.pdf} 
%     \caption{Amazon Web Services environment}
%     \label{fig:fig-3}
% \end{figure*}

%%%%%%%%%%%%%%%%%%%%%%%%%%%%%%%%%%%%%%%%%%%%%%%%%%%%%%%%

\begin{itemize}
    \item \textbf{AWS Documentation:} AWS documentation provides a comprehensive resource that explains the features and best practices of each service.

    \item  \textbf{Cloud Computing Service Comparison:} Analyses AWS alongside other cloud platforms, including key services such as Elastic Compute Cloud (\textbf{EC2}), Simple Storage Service (\textbf{S3}), and \textbf{Lambda}.
    
    \item \textbf{Big Data and Analytics in AWS:} Big Data and Analytics in AWS examines the effectiveness of AWS's big data tools, such as Elastic MapReduce (\textbf{EMR}) and \textbf{Kinesis}, for large-scale data handling.

    \item \textbf{Machine Learning on AWS:} Machine Learning on AWS discusses \textbf{SageMaker} and other AI/ML services, with a focus on deployment and scalability.
    
    \item \textbf{AWS Security and Compliance:} Security and Compliance of AWS examines how AWS security features such as Virtual Private Cloud (\textbf{VPC}), Identity and Access Management (\textbf{IAM}), and \textbf{CloudFront} can help ensure data protection and regulatory compliance.
    
\end{itemize}

%%%%%%%%%%%%%%%%%%%%%%%%%%%%%%%%%%%%%%%%%%%%%%%%%%%%%%%%
%\vspace{-10pt}

AWS capabilities are highlighted, as are the various ways in which organisations can benefit from its comprehensive cloud solution.

%%%%%%%%%%%%%%%%%%%%%%%%%%%%%%%%%%%%%%%%%%%%%%%%%%%%%%%%
\subsection{Microsoft Azure environment}

Here are some useful tools for exploring Microsoft Azure's cloud services in computing, storage, data analytics, machine learning, networking, security, and developer tools\footnote{https://docs.microsoft.com/en-us/azure}:

%%%%%%%%%%%%%%%%%%%%%%%%%%%%%%%%%%%%%%%%%%%%%%%%%%%%%%%%

%\begin{figure}[ptbh] 
%\includegraphics[scale=0.38]{ACM-fig-4.pdf}
%\caption{} \label{fig:fig-4}
%\end{figure}

%\vspace{-5pt} % Before the figure
% \begin{figure*}[!t]
%     \centering
%    %\vspace{-10pt} % Adjust this value as necessary
%     \includegraphics[width=1\textwidth]{AZURE.pdf} % Adjust width as needed
%     %\captionsetup{ width=0.7\textwidth} % Shift caption left
%     \caption{Microsoft Azure environment}
%     \label{fig:fig-4}
% \end{figure*}

%%%%%%%%%%%%%%%%%%%%%%%%%%%%%%%%%%%%%%%%%%%%%%%%%%%%%%%

\begin{itemize}
    \item \textbf{Azure Documentation:} Every service, including compute, storage, data, and security capabilities, is explained in detail in Microsoft's extensive documentation on Azure.
    
	\item \textbf{Comparative Analysis of Cloud Services:} This article highlights Microsoft Azure's adaptability in business settings while comparing its offerings with those of other top cloud platforms.
    
	\item \textbf{Azure Data Analytics and Big Data Services:} The roles of \textbf{Azure Synapse}, \textbf{HDInsight}, and Data Lake in big data analytics are discussed, as well as their applications.
    
	\item \textbf{Azure Machine Learning:} This study investigates the ways in which automated ML workflows and sophisticated AI applications are supported by Azure Machine Learning and Cognitive Services.
    
	\item \textbf{Azure Security:} The security features of Azure, such as Active Directory, Security Centre, and Key Vault, are examined in this article, along with how they handle enterprise security issues.
    
	\item \textbf{Monitoring and Management in Azure:} Covers Azure Monitor, Application Insights, and Log Analytics, with a focus on resource monitoring and management.
    
	\item \textbf{Azure Developer Tools:} Azure developer tools highlight the role of Azure Development Operations (\textbf{DevOps}), \textbf{GitHub Actions}, and \textbf{Visual Studio App Centre} to support development and continuous integration workflows.
    
\end{itemize}

%%%%%%%%%%%%%%%%%%%%%%%%%%%%%%%%%%%%%%%%%%%%%%%%%%%%%%%%
%\vspace{5pt} % Before the figure
These sources give Azure's wide range of cloud services and capabilities more context and support for a range of business requirements.

%%%%%%%%%%%%%%%%%%%%%%%%%%%%%%%%%%%%%%%%%%%%%%%%%%%%%%%%

%\begin{figure}[ptbh] 
%\includegraphics[scale=0.5]{ACM-fig-5.pdf}
%\caption{} \label{fig:fig-5}
%\end{figure}

%\vspace{-5pt} % Before the figure
% \begin{figure*}[!htbp]
%     \centering
%    %\vspace{-10pt} % Adjust this value as necessary
%     \includegraphics[width=1\textwidth]{ICDM-fig-5.pdf} % Adjust width as needed
%     %\captionsetup{ width=0.7\textwidth} % Shift caption left
%     \caption{Mind-map of AWS, Azure, and Google Cloud platforms.}
%     \label{fig:fig-5}
% \end{figure*}
%%%%%%%%%%%%%%%%%%%%%%%%%%%%%%%%%%%%%%%%%%%
%\vspace{5pt}

The main characteristics, classifications, and particular services provided by various cloud platforms (like AWS, Azure, and Google Cloud) are depicted clearly in a mind map\footnote{\url{https://github.com/GenomicAI/BigData/blob/main/mindmap.pdf}}, which offers a visual representation of the tools available within each cloud environment. To help users quickly understand the depth and capabilities across multiple functional areas, including Compute Services, Storage Solutions, Networking, Data Analytics, Machine Learning and AI, Security and Compliance, and Developer Tools, each cloud provider's tools are arranged into branches within the mind map. An overview of each cloud environment is included with the mind map, providing a concise synopsis of each platform's key features and distinctive advantages. Users can make well-informed decisions about which cloud provider and services best suit their unique project needs with the help of this high-level overview.

%%%%%%%%%%%%%%%%%%%%%%%%%%%%%%%%%%%%%%%%%%%%%%%%%%%%%%%%
%\vspace{-8pt}
\section{ Recent Advancement in Big Data Architecture}

\textbf{GenAI}

GenAI a transformative force in big data architecture, holds the potential to revolutionise data generation, pipeline automation, storage optimisation, real-time analytics, and security. AI-driven models such as Generative Adversarial Networks (GANs) and Variational Autoencoders (VAEs) are paving the way for synthetic data generation, enhancing data availability and model robustness while ensuring privacy compliance \cite{balasubramaniam2024road}. GenAI's capability to streamline data pipeline automation and enhance workflow efficiency presents a promising avenue for further research. Despite these benefits, GenAI still faces notable challenges, including substantial computational costs, ethical dilemmas, and the complexity of AI interpretability. Addressing these concerns requires a focus on developing resource-efficient AI models and advancing methodologies for explainable AI in future research. This will not only improve trust and scalability in AI-driven big data architectures but also invoke a sense of determination in our collective efforts to overcome these challenges.

\textbf{DeepSeek} 

DeepSeek is revolutionising big data architecture by developing highly efficient AI models that significantly reduce computational costs while maintaining high performance \cite{guo2024deepseek}. This efficiency stems from optimised data handling and reduced hardware dependency, making AI more accessible and scalable. By using an open-source approach, DeepSeek fosters AI transformation, enabling businesses and researchers to develop AI solutions without relying on expensive infrastructure. 
However, challenges remain, particularly in regulatory oversight and ethical concerns, as the widespread availability of powerful AI models raises concerns about misuse and accountability. Despite these risks, DeepSeek's innovations offer a promising shift towards more cost-effective and widely accessible AI-driven big data architectures.

%%%%%%%%%%%%%%%%%%%%%%%%%%%%

\begin{figure*}[!th]
    \centering
   %\vspace{-10pt} % Adjust this value as necessary
    \includegraphics[width=1\textwidth]{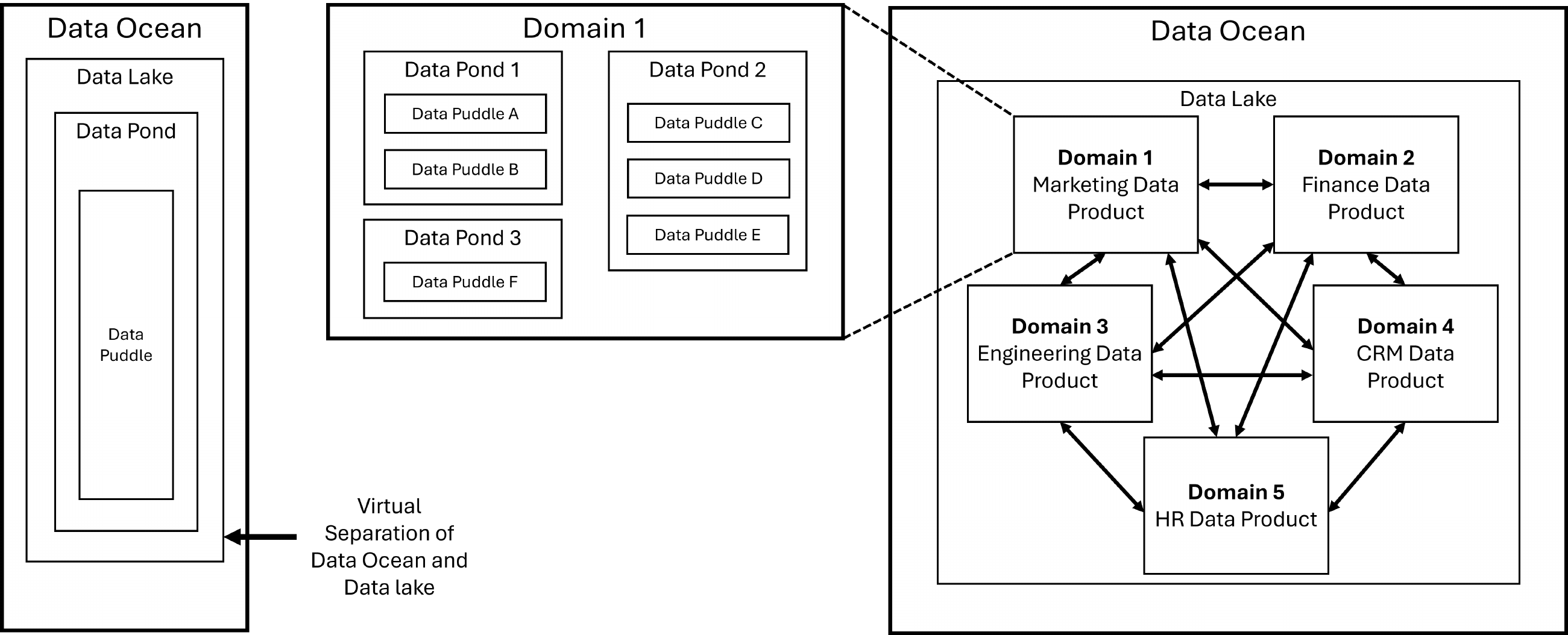} % Adjust width as needed
    %\captionsetup{ width=0.7\textwidth} % Shift caption left
    \caption{In Data Mesh Architectures the individual functions(Marketing/Finance/etc) behave as separate domains having the ownership of their data. Data is considered as products and is distributed among the domains/owners. The simple premise of data mesh is that business domains should be able to define, access, and control their own data products.}
    \label{fig:fig-6}
\end{figure*}

\textbf{Data mesh} 

Data mesh is a modern architectural paradigm that promotes decentralised data management by allocating data ownership to specific business domains. This model enables the creation, maintenance, governance, and distribution of data products, ensuring they are easily accessible and reusable by various data consumers. Organisations can achieve greater scalability and agility by shifting away from centralised data architectures and optimising their data workflows to support real-time analytics and cross-functional collaboration better \cite{dibouliya2023review}.

\textbf{Data Ocean}

The Data Ocean concept represents (Figure~\ref{fig:fig-6}) the most advanced stage in the evolution of data storage and management, encompassing all enterprise data in its raw, unprocessed form while encouraging self-service access and data-driven decision-making throughout the organisation. The first step in this evolution is a Data Puddle, which is a small-scale, one-off data mart that uses big data technologies to meet particular project needs. Multiple puddles combine to form a Data Pond, a larger repository that unifies disparate isolated datasets or acts as an offload from conventional data warehouses as data needs grow. The following step, Data Lake, centralises a large amount of data that might or might not be immediately useful for business purposes. This enables users to independently explore and extract insights without the need for IT support. The most complete stage is the Data Ocean, which unifies raw data from all business domains, facilitates cross-functional cooperation and promotes an enterprise-wide data mesh strategy. This ensures accessibility, governance, and scalability for data-driven innovation by enabling business domains such as marketing and finance to own and manage their data as products.

%%%%%%%%%%%%%%%%%%%%%%%%%%%%

\section{ Challenges and Future Work}

Big data cloud architecture has progressed rapidly in the last decade with advanced tools and processes, but there are various challenges that need to be addressed. These challenges include being outdated in several tools and concepts, which require updating from time to time. Also, this blueprint architecture is only a reference architecture to organisations, and many organisations do not require all the layers and tools mentioned in the big data architecture. In this instance, they need to assess the individual requirements and scalability in carefully selecting tools. Also, the cost consideration has not been considered in the representation of this blueprint. This requires thorough analysis in selecting the correct cloud environment and tools. In future work, we propose to have detailed papers on each tool and its usability for the user's convenience in selecting the right tool for their business requirements. Possible cloud comparisons for similar types of tools will be an added advantage for business users who are using them practically.

%%%%%%%%%%%%%%%%%%%%%%%%%%%%%%%%%%%%%%%%%%%%%%%%%%%%%%%%
%\vspace{-10pt}

\section{Conclusion}

Big data has changed the way businesses build their technological infrastructures and make decisions. Businesses require a strategic and well-structured architecture in order to use this huge and diverse technology stack. Data ingestion, storage, processing, and analysis are among the fundamental data processes that are addressed by the suggested blueprint architecture framework, which also guarantees faithfulness to important governance, security, and regulatory standards.

The novelty of this study lies in its practical approach to implementing a standardized big data functional architecture across multiple cloud platforms—Google Cloud, AWS, and Microsoft Azure. Unlike existing studies that focus on theoretical models or platform-specific solutions, this research provides a comparative and adaptable framework, enabling organizations to select or modify their implementation based on cost efficiency, scalability, and business needs. By offering a structured pathway for deploying big data solutions in a cloud-agnostic manner, this study empowers businesses to make informed decisions while optimizing their technological investments.

Additionally, this study incorporates emerging trends such as Generative AI (GenAI) and low-code machine learning, which are revolutionizing data-driven decision-making and automation. These advancements allow organizations to accelerate data processing and analytics while minimizing the complexity of AI adoption. By integrating these cutting-edge technologies into the big data architecture, this research ensures that businesses remain at the forefront of innovation, leveraging the latest capabilities to enhance efficiency, security, and strategic flexibility.

This blueprint, designed with scalability and integration in mind, helps businesses build adaptable and robust data strategies. By bridging the gap between theoretical models and real-world implementation, this research provides a practical, forward-looking foundation for enterprises to harness big data effectively and cost-efficiently across diverse cloud environments.

\bibliographystyle{ieeetr}
\bibliography{main.bib}

\end{document}